# Operation of Quantum Plasmonic Metasurfaces Using Electron Transport through Subnanometer Gaps

*Takashi Takeuchi\*, Masashi Noda, and Kazuhiro Yabana*

Center for Computational Sciences, University of Tsukuba, Tsukuba 305–8577, Japan

ABSTRACT: Herein, we investigate the optical properties of quantum plasmonic metasurfaces composed of metallic nano-objects with subnanometer gaps according to the time-dependent density functional theory, a fully quantum mechanical approach. When the quantum and classical descriptions are compared, the transmission, reflection, and absorption rates of the metasurface exhibit substantial differences at shorter gap distances. The differences are caused by electron transport through the gaps of the nano-objects. The electron transport has profound influences for gap distances of ≲ 0.2 nm; that is, approximately equal to half of the distance found in conventional gap plasmonics in isolated systems, such as metallic nanodimers. Furthermore, it is shown that the electron transport makes the plasmon features of the metasurface unclear and produces broad spectral structures in the optical responses. In particular, the reflection response exhibits rapid attenuation as the gap distance decreases, while the absorption response extends over a wide spectral range.

A plasmonic metasurface within which metallic nano-objects are (a) smaller than the wavelength of the incident light and (b) periodically placed on a plane, has been attracting substantial attention in terms of its highly beneficial optical characteristics.[1–2] Subject to the irradiation of an optical field, the electromagnetic energy of light is confined in the constituent nano-objects as a plasmon—a collective motion of electrons—generating locally enhanced plasmonic fields around the objects. This plasmon property, manifested by its enhanced optical capturing capacity, has opened up many practical applications of the metasurface, such as subdiffraction lensing,[3–4] monochromatic or color holography,[5–8] polarization converters,[9–10] polarization-selective elements,[11–12] and others.[13–14]

In most investigations conducted to date, the optical properties of metasurfaces that are composed of nano-objects whose gap distances are at the wavelength or subwavelength scales have been explored. Recently, experimental studies have reported on metasurfaces with gap distances as small as a subnanometer.[15–16] For example, the optical properties of a plasmonic metasurface with subnanometer gaps made of gold nanospheres coated with tunable alkanethiol ligand shells were reported[16] in which the interparticle distance was controlled from 0.45 to 2.8 nm. It has been found that owing to the significantly intense optical confinement at the gaps, the metasurface yielded a high-refractive index that could not be achieved in naturally occurring transparent materials.

In the last decade, tremendous attention had been paid to subnanometer gap plasmonics in isolated systems with metallic nanodimers constituting a typical paradigm.[17–23] In brief, the research findings were based on interesting observations in which substantial differences existed between descriptions using classical and quantum theories. In the classical description, in which the surface of nanoparticles are modeled as 'discontinuous' boundaries, the solutions of Maxwell's equations yield an extremely enhanced and confined electromagnetic field at the gap as the nanoparticles come closer to each other. By contrast, in quantum description with 'continuous' boundaries expressed using realistic electron density distributions, the charge transfer processes across the gap that were caused by electron tunneling suppressed the field enhancements, especially at small gap distances ($\lesssim$ 0.4 nm).[17, 19, 21–22, 24] This suppression in the gap plasmonics has been confirmed by experiments.[20, 25–26] It is often crucially important to know when the quantum effect starts to appear; for example, in plasmonic applications that are based on strong optical enhancement and confinement, including in surface-enhanced Raman scattering (SERS) technologies[27–30] and other nanophotonic systems.[31–34]

In view of the broad applicability of metasurfaces, it is particularly concerning how the quantum effects appear in them. However, to the best of our knowledge, there have been no prior experimental or theoretical reports that have discussed the quantum effects. Measurements carried out on the metasurface with a minimum gap distance of 0.45 nm[16] have been reproduced by classical theory. This may be reasonable owing to fact that the effects of electron tunneling on isolated systems with gap distances of $\gtrsim$ 0.4 nm have been small.

In this study, quantum plasmonic metasurfaces composed of metallic nanospheres with subnanometer gaps were theoretically investigated by a fully quantum mechanical method using the time-dependent density functional theory (TDDFT)[35–36] combined with a two-dimensional (2D) course-graining approach.[37] We will show the transmission, reflection, and absorption rates of the metasurface to elucidate the quantum effects in the optical properties. Our results indicate that there are substantial differences between the classical and the quantum descriptions for metasurfaces with gap distances $\lesssim$ 0.2 nm. This gap distance is almost equal to half of the critical threshold distance in isolated systems presented in previous studies.[17, 19–22, 25-26] Our results also indicate that the quantum effects make the plasmonic features unclear in the frequency domain and produce broad spectral structures in the optical responses at small gap distances. In particular, the reflection shows rapid attenuation as the gap distance decreases, while the absorption extends over a broad frequency range. These findings indicate that they are suitable as optical absorbers.

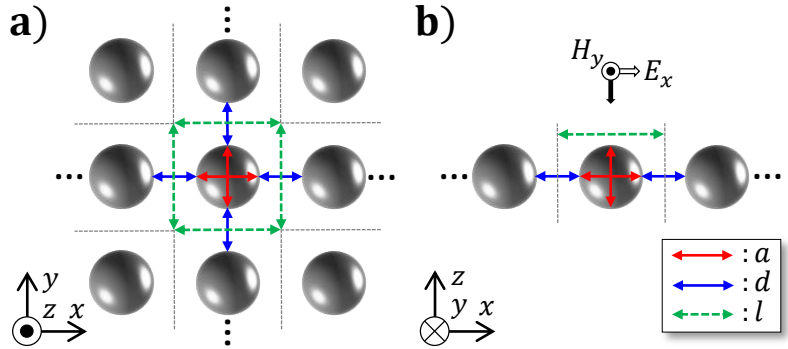

**Figure 1.** Schematics of the studied metasurface consisting of nanospheres periodically arrayed on the *xy* plane: (a) top and (b) side views. The incident light has $E_x$ and $H_y$ components and propagates toward the negative *z* direction. The symbols *a*, *d*, and *l*, drawn using red, blue, and green arrows, denote the diameters of the spheres, the gap distances, and the lengths of the periods, respectively.

The studied system is displayed in Figure 1a and 1b. In these figures, metallic nanospheres with diameter *a* are periodically arrayed on the *xy* plane with a gap distance *d* and a period length *l*. The incident light is linearly polarized with the $E_x$ and $H_y$ components, and propagates along the negative *z* direction. To take into account quantum mechanical effects with a moderate computational cost, we employed the spherical jellium model (JM) that ignores ionic structures. Despite its simplicity, the JM can describe the actual plasmonic behavior of electrons in metallic nanoparticles.[17, 19, 38] In particular, qualitative agreements with measurements have been reported for an isolated metallic nanodimer with a subnanometer gap that accounted for the effects caused by electron tunneling.[20, 25]

In the JM, the Wigner–Seitz radius $r_s$ related to the average charge density, *n*, $n^+ = ((4\pi)r_s^3/3)^{-1}$, specifies the medium. Herein, $r_s = 4.01$ Bohr, which corresponds to the Na metal. Each nanosphere was set to accommodate 398 electrons that correspond to a closed shell structure. The resultant diameter of the nanosphere $a$ was 3.1 nm. The size is sufficiently large so that the nanoparticle exhibits a typical plasmonic resonance.[21]

To calculate the optical responses of metasurfaces, we employed linear-response TDDFT, which is extensively used for the optical responses of molecules[39] and solids[40] at a first-principles level. We assumed an adiabatic approximation for the exchange-correlation potential and used the local density approximation.[41] Numerical calculations were carried out using SALMON, an open-source code (https://salmon-tddft.jp/) developed in our group.[42] The Supporting Information contains a detailed description of the adopted numerical approach.

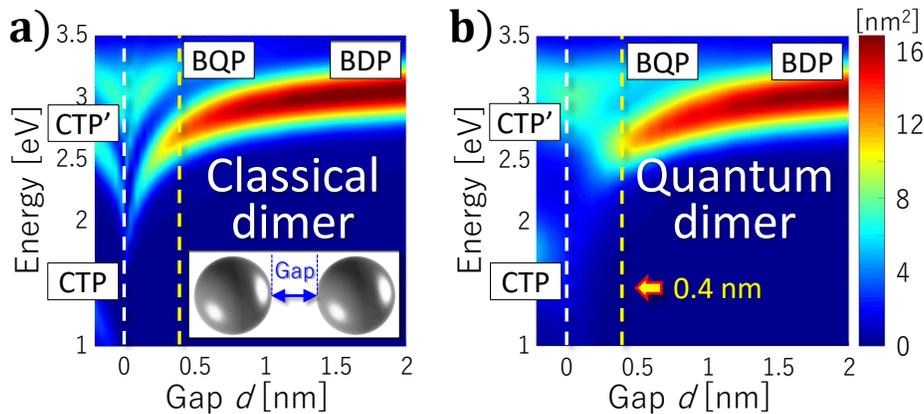

**Figure 2.** (a) Optical absorption cross-section of a metallic dimer (the inset shows the relevant schematic) calculated based on classical theory, whereby the constituent nanospheres have the same diameters as those in the case of the metasurface, as shown in Figure 1. The horizontal and vertical axes denote the gap distance $d$ and optical frequency, respectively. The white dashed lines indicate the loci at which $d$ = 0 nm, while the yellow lines indicates loci at which $d$ = 0.4 nm. (b) Results generated by the TDDFT in conjunction with the JM.

First, to facilitate a comparison between the isolated and periodic systems, we briefly describe the optical response of a nanodimer that consists of metallic nanospheres identical to those used in the metasurface calculations. Figure 2a and 2b show the absorption cross-sections of the dimer calculated using the classical theory and the TDDFT, respectively. In the classical theory, the Drude model was used. Details of the theories are explained in the Supporting Information. In both theories, the absorption properties are dominated by the bonding dipolar plasmon (BDP) mode for gap distances $d \gtrsim 0.4$ nm. The red shift that appears in both theories is caused by the increase in the coupling strength as the two spheres approach each other. However, when $d \lesssim 0.4$ nm, substantial differences can be observed between the two results. In the classical theory, there appears an additional bonding quadrupolar plasmon (BQP) mode at a higher frequency region until $d = 0$ at the contact point of the two spheres. For $d < 0$ nm after the geometrical overlap of the spheres, the charge transfer and its high-order-plasmon (CTP and CTP') modes emerge. By contrast, in the TDDFT, quite different spectral distributions are manifested for $d \lesssim 0.4$ nm. The BDP disappears prior to the direct contact and a hybridized form of the BQP and the CTP becomes visible. This is caused by the charge transfer due to the electron tunneling through the gap region between the nanospheres. These trends had already been established in previous studies.[17, 19–22]

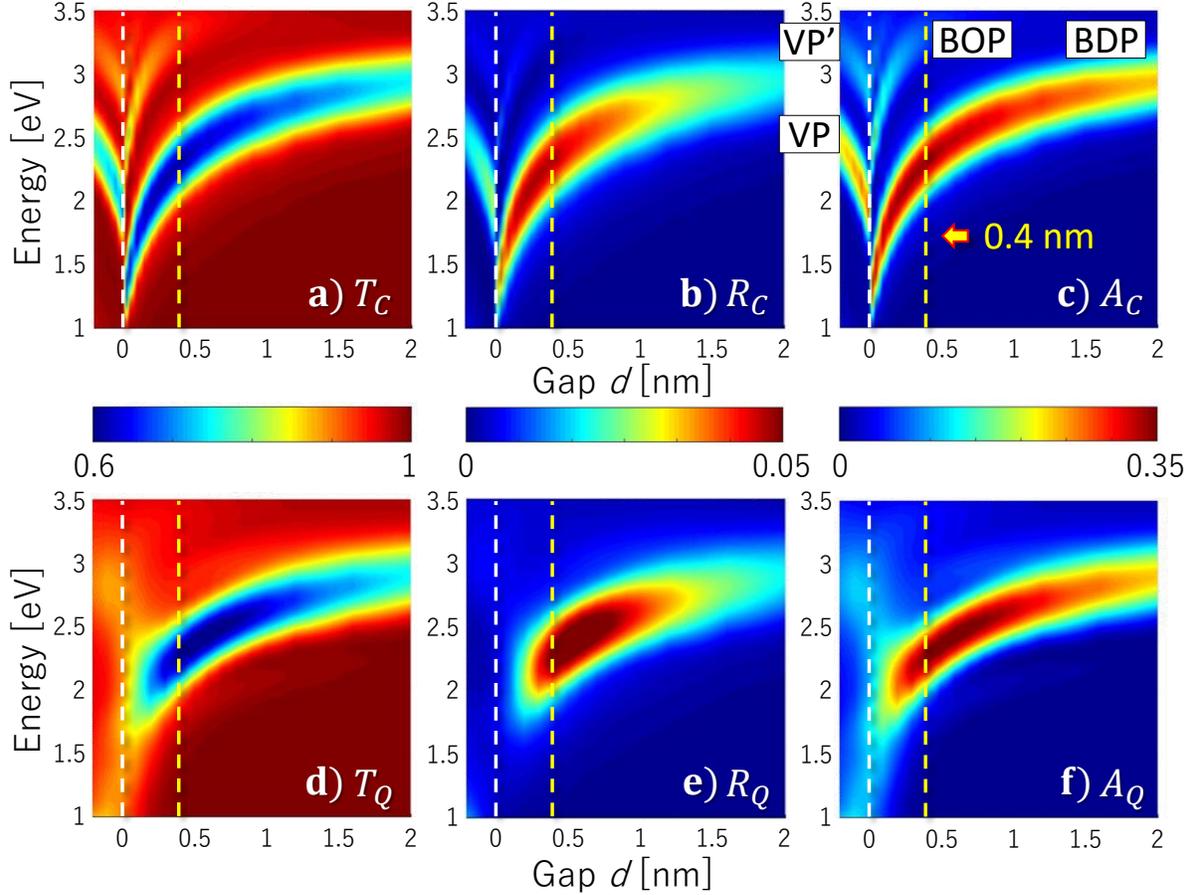

**Figure 3.** Spectral distributions of the transmission (*T*), reflection (*R*), and absorption (*A*) rates of the metasurfaces calculated by the classical theory (a–c; subscript *C*) and by the TDDFT (d–f; subscript *Q*). A common color scale applies for *C* and *Q*. All distributions are shown as functions of the gap distance *d* and the optical frequency. Similar to Figure 2a and 2b, the white and yellow dashed lines indicate the loci at which $d = 0$ and 0.4 nm, respectively.

We now move to our main subject of the present study. Figure 3 shows the calculated spectral distributions of the transmission (*T*), reflection (*R*), and the absorption rates (*A*) of the plasmonic metasurfaces for gap distances ranging from $d = -0.2–2$ nm. They are obtained by employing a 2D coarse graining approach.[37] The relevant details are described in the Supporting Information. Figure 3a–3c shows results based on the use of the classical theory, while Figure 3d–3f shows the results estimated based on the TDDFT with the use of the JM. For $d \gtrsim 0.4$ nm, the quantum and classical descriptions generate almost similar results that are characterized by a single peak derived from the BDP mode. Even though the BDP mode also appears in the isolated system, as shown in Figure 2, we find much larger red shifts compared with the nanodimer case. This is rationalized by the fact that the constituent nanospheres in the metasurface suffer a more intense electromagnetic coupling effect compared with the dimer case owing to its periodically arrayed geometry. These features match well with those reported in the previous investigation,[16] whereby the plasmonic gap distance was explored to be as small as 0.45 nm in the measurements and simulations according to the classical theory. When $d \lesssim 0.4$ nm, the bonding octopolar plasmon (BOP) mode emerges, and is clearly visible in the classical description at a higher frequency region. Compared with the isolated system, the BQP mode does not appear in the optical response because of the full symmetry on the *xy* plane. By decreasing the distance to $d = 0$ nm or less, the void plasmon (VP) and its higher-order plasmon (VP') modes appear when the classical theory is used.

We now look into the electron transport effects at small *d* regions. As has been depicted in Figure 2c and 2d, the BDP mode disappears at $d = 0.4$ nm in the case of the nanodimer. Conversely, as observed in Figure 3d–3f, the BDP mode excited on the metasurface survives within a much smaller distance, which can be as small as $d \approx 0.2$ nm. As it has already been noted, the resonant frequency of the BDP in the metasurface is much smaller than that in the nanodimer. Thus, the changes in the optical responses caused by the classical and quantum effects are different between the nanodimer and the metasurface, reflecting their geometrical differences. These results indicate that the studied metasurface maintains a strong optical confinement and enhancement in the gap distance as small as $d \approx 0.2$ nm, which is much smaller than the critical distance $d \approx 0.4$ nm in the nanodimer. Furthermore, although the classical theory yields similar spectral distributions among *T*, *R*, and *A*, those calculated by the TDDFT yield a difference at $d \lesssim 0.2$ nm, *R* related to the BDP vanishes immediately, while *T* and *A* remain and are extensively spread in frequency, representing unclear plasmonic features of the constituent nanoshperes in the metasurface. In particular at $d =$ 0 nm, R is less than 0.5%, while A is maintained at 10% in the visible frequency range. This indicates that they are suitable for an optical absorber.[43–46]

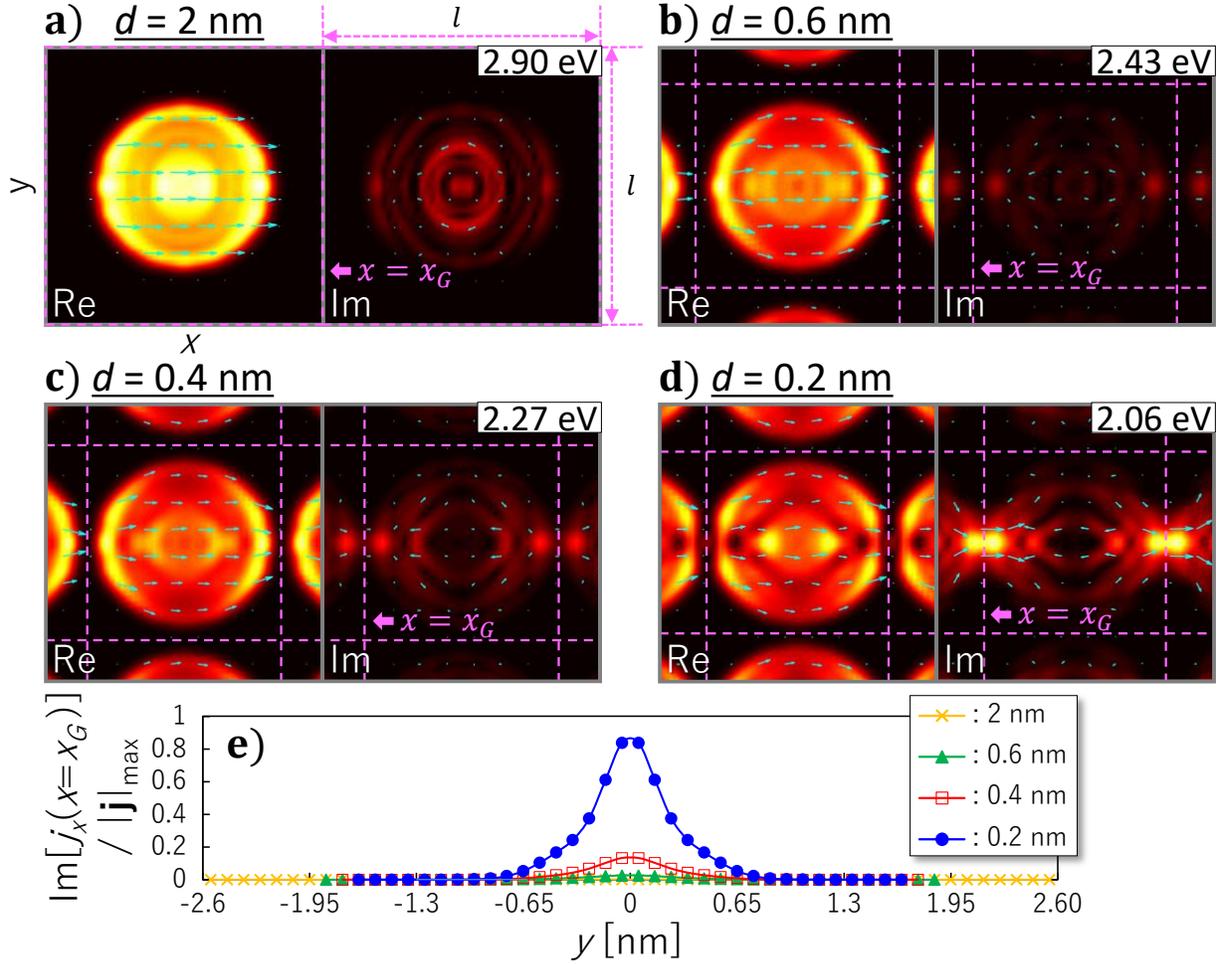

**Figure 4.** (a–d) Spatial distributions of electric current density on the $xy$ plane ($z = 0$) at the resonant frequency of the bonding dipolar plasmon (BDP) mode, $\omega_{BDP}$ (shown in the top-right white boxes). The maximum intensity is normalized in each figure, showing $\mathbf{j}(x, y, z = 0, \omega = \omega_{BDP})\,|/|\,\mathbf{j}\,|_{max}$. Real- and imaginary-parts are shown in the left and the right panels, respectively, for $d = 2, 0.6, 0.4$, and $0.2$ nm. The light blue arrows indicate that their vector distributions with their lengths are associated with the magnitude. Dashed pink lines indicate the unit cell with a period $l$. The electron transport takes place through the $x = x_G$ plane. (e) The imaginary parts of the currents are displayed as a function of $y$ at the plane $x = x_G$.

To elucidate the mechanism that produces the differences between the classical and quantum descriptions, we visualized the spatial distributions of the electric current density, $\mathbf{j}(\mathbf{r}, \omega)$, calculated by the TDDFT. For an electric field with frequency omega, $\mathbf{E}(t) = \mathrm{E}_0\cos(\omega t)$, the electric current density is given by $\mathbf{j}(\mathbf{r}, t) = \text{Re}[\mathbf{j}(\mathbf{r}, \omega)]\cos(\omega t) - \text{Im}[\mathbf{j}(\mathbf{r}, \omega)]\sin(\omega t)$. We refer to $\text{Re}[\mathbf{j}(\mathbf{r}, \omega)]$ as the conductive current because it has the same phase relation to the applied electric field. Additionally, we refer to $\text{Im}[\mathbf{j}(\mathbf{r}, \omega)]$ as the dielectric because it has a phase difference of $\pi/2$. Similar investigations by the classical theory are presented in the Supporting Information. Figure 4a–4d shows the electric current density in the $xy$ plane ($z = 0$) at the resonant frequencies of the BDP mode, $\omega_{BDP}$. The maximum value is normalized in each figure, showing $\mathbf{j}(x, y, z = 0, \omega = \omega_{BDP})$ $|/|\mathbf{j}|_{max}$. The real and imaginary parts of the current are shown in the left and right panels, respectively, for $d = 2, 0.6, 0.4$, and $0.2$ nm. The light blue arrows indicate the vector distributions of the current, whereby their lengths are associated with the magnitudes. The dashed pink lines indicate the unit cell with a period $l$. At $d = 2$ nm, $\text{Re}[\mathbf{j}]$ is almost spatially uniform inside the sphere, as observed in Figure 4a. As $d$ decreases, the currents tend to be localized at both ends of the sphere. This localization, which causes the red shift of the BDP mode as observed in Figure 3, is caused by the attractions among the positive and negative charges that appear at both ends of the neighboring nanospheres.[17, 19–22] As it has been mentioned in the Supporting Information, this behavior is also observed in the classical description. By contrast, as shown in Figure 4d, at $d = 0.2$ nm, the current distribution in the TDDFT shows a drastic change that is not observed in the classical theory. In $\text{Re}[\mathbf{j}]$, there appears a hole in the electric current distribution at both ends of the sphere along the $y = 0$ axis. In $\text{Im}[\mathbf{j}]$, a clear opening of the current owing to the electron transport is observed in the $x = x_G$ plane along the $y$ axis. To elucidate the electron transport clearly, we plot the current distribution $\text{Im}[\mathbf{j}]$ against $y$ in the $x = x_G$ plane in Figure 4e using the gap distances of Figure 4a–4d. This shows that the electron transport increases gradually as $d$ decreases up to around 0.4 nm, and then increases abruptly toward $d = 0.2$ nm. The $\mathbf{j}/|\mathbf{j}|_{max}$ value at $d = 0.2$ nm is close to 0.9, indicating that the current is almost maximized at the gap plane $x_G$. This abrupt

change of the current can be rationalized by the change of the relation between the potential barrier height and the energy of the highest occupied orbital. In fact, we have found that they coincide with each other at around $d = 0.3$nm. Therefore, the electron transport process can be classified to the two situations: (a) At $d \gtrsim 0.3$ nm, electron transport takes place by the quantum tunneling; (b) at d $\lesssim$ 0.3 nm, electron transport is dominantly caused by the over-barrier process so that a large amount of electrons can move through the gaps. The electron transport over the potential barrier rationalizes the disappearance of the BDP mode at $d = 0.2$nm, as shown in Figure 3d–3f, and displays the qualitative differences in the results between the quantum and the classical descriptions.

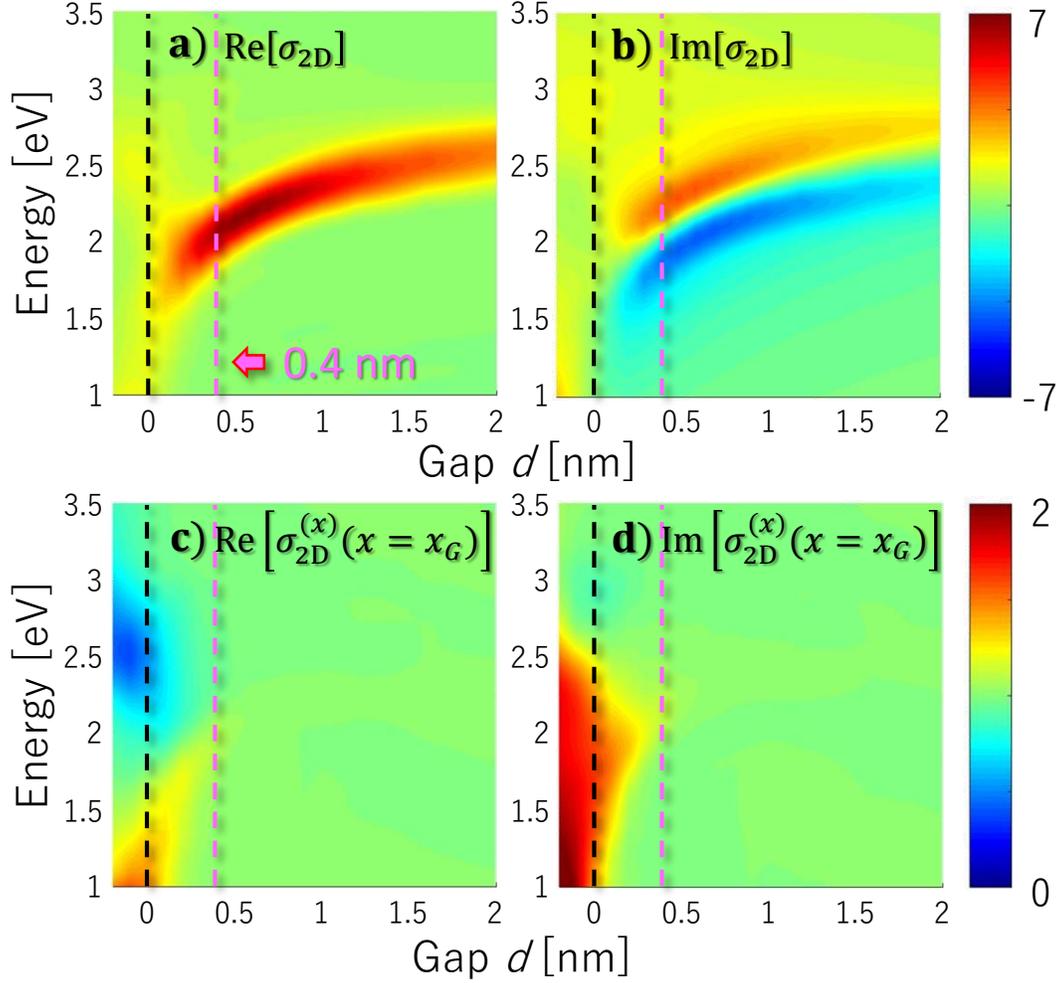

**Figure 5.** Spectral distributions of the two-dimensional (2D) complex conductivities of the metasurface as a function of the gap distance $d$ (Re[$\sigma_{2D}$] is plotted in (a) and Im[$\sigma_{2D}$] in (b) in atomic units). The contributions to the 2D complex conductivity at the gap plane $x = x_G$ are shown [Re[$\sigma_{2D}^{(x)}(x = x_G)$] in (c) and Im[$\sigma_{2D}^{(x)}(x = x_G)$] in (d)]. Precise definitions are described in the Supporting Information. A common color scale applies to the real and the imaginary parts. The black and pink dashed lines indicate the loci at which $d = 0$ nm and 0.4 nm, respectively.

To clarify the effects of electron transport within the entire spectral region, we show the spectral distributions of the 2D complex conductivity of the metasurface, with Re[$\sigma_{2D}$] and Im[$\sigma_{2D}$] in atomic units in Figure 5a and 5b, respectively. The optical quantities $T$, $R$, and $A$, in Figure 3d–3f are obtained from the 2D conductivity. As explained in the Supporting Information, it is possible to express the 2D conductivity as an integration over $x$, $\sigma_{2D}(\omega) = (\frac{1}{l})\int dx \sigma_{2D}^{(x)}(x, \omega)$. Figure 5c and 5d show a contribution at the gap plane, $x = x_G$, to the conductivity, $\text{Re}[\sigma_{2D}^{(x)}(x = x_G)]$ and $\text{Im}[\sigma_{2D}^{(x)}(x = x_G)]$. In the region $0 < d < 0.2$ nm, a negative contribution (blue-colored region) appears in $\text{Re}[\sigma_{2D}^{(x)}(x = x_G)]$. This is caused by the current density of the VP mode whose direction is reversed around the gap plane $x = x_G$. In $\text{Im}[\sigma_{2D}^{(x)}(x = x_G)]$ in $d < 0.2$ nm, a positive contribution appears in the broad frequency region. As observed from Figure 5c and 5d, which displays $\text{Re}[\sigma_{2D}^{(x)}(x = x_G)]$ and $\text{Im}[\sigma_{2D}^{(x)}(x = x_G)]$, the electron transport at the gap plane $x = x_G$ contributes substantially to the 2D conductivity. Based on this result, we conclude that the onset of electron transport is at approximately $d \approx 0.2$ nm.

The execution of this study was based on various approximations, which are outlined below in conjunction with the limitations of this study. Although the JM is known to provide a reasonable description for plasmonic features of metallic nanoparticles, descriptions of the tunneling effect between nanoparticles may not be realistic enough because the tunneling is sensitive to a precise structure at the interfaces at the atomic scale, as has been shown in recent studies.[21] Electron tunneling was studied[21] in a metallic nanodimer consisting of realistic Na clusters. This study revealed that the JM could not account for the lightning rod effect that is caused by the electron concentration at the edges of the clusters. First-principles calculations that incorporate precise ionic structures could account for these effects in our targeting metasurface. However, in practice, this is a rather difficult task because the shapes of the fabricated nanoparticles arrayed on the metasurface are quite random.[15–16] Accordingly, we could not apply the periodic boundary condition that was used in the present work.

Although we have considered very small nanoparticles whose diameters are 3.1 nm, insights into the electron transport should have robustness for larger systems because previous studies that dealt with nanodimers showed substantial agreements despite the spatial scale differences between theory and measurements.[21] To explore the electron tunneling effects, the quantum corrected model (QCM)[19] was developed and successfully applied to isolated systems, such as nanodimers. This model is based on the classical theory with some modifications that take into account the electron tunneling effects. While the TDDFT has been used in this study, the QCM is also a prospective candidate for the exploration of quantum plasmonic metasurfaces with subnanometer gaps. Furthermore, given that the QCM requires a moderate computational cost compared with the expensive TDDFT, it will be useful for investigating larger systems. We consider that the application of the QCM to the metallic metasurface and the comparison with the TDDFT results is an important topic for a future study.

In conclusion, this study has presented a theoretical investigation into quantum plasmonic metasurfaces with subnanometer gaps. The optical responses of a metasurface composed of two-dimensionally arrayed metallic nanospheres have been examined based on the use of the TDDFT with the JM. We have shown the transmission, reflection, and absorption rates of the metasurface as a function of the gap distance. It has been shown that substantial differences exist between the classical and the quantum theories when the gap distances become less than approximately 0.2 nm owing to the electron transport. This threshold distance is almost half of the typical threshold distance known for isolated systems, such as metallic nanodimers. It has also been shown that the quantum effects make the plasmon features unclear and produce broad spectral structures in the optical responses. In particular, the reflection shows rapid attenuation as the gap distance decreases, while the absorption response is maintained over a broader spectral range. The knowledge obtained in this study is expected to be beneficial in view of the growing interests in metasurfaces with gap plasmonics, and will provide relevant guidelines for their design.

## ASSOCIATED CONTENT

**Supporting Information**

Detailed theoretical descriptions of 2D coarse graining approach, the TDDFT, and classical theory with plasmonic responses are given. This material is submitted as a separate file.

## AUTHOR INFORMATION

**Corresponding Author**

*E-mail: take@ccs.tsukuba.ac.jp

**Author Contributions**

**Funding Sources**

This research was supported by the JST–CREST under Grant No. JP–MJCR16N5, and by the JSPS KAKENHI under Grant No. 15H03674.

## ACKNOWLEDGMENT

Theoretical computations were performed in part with the use of the computational resources of Oakforest–PACS provided by the Multidisciplinary Cooperative Research Program in Center for Computational Sciences, University of Tsukuba, and the Research Center for Computational Science, Okazaki, Japan.

# Supporting information

Operation of Quantum Plasmonic Metasurfaces Using Electron Transport through Subnanometer Gaps

**1. Theory on the optical responses of metasurfaces**

We summarize the theory applicable to the optical responses of two-dimensional (2D) materials utilized for metasurfaces based on the adoption of a 2D course graining approximation.

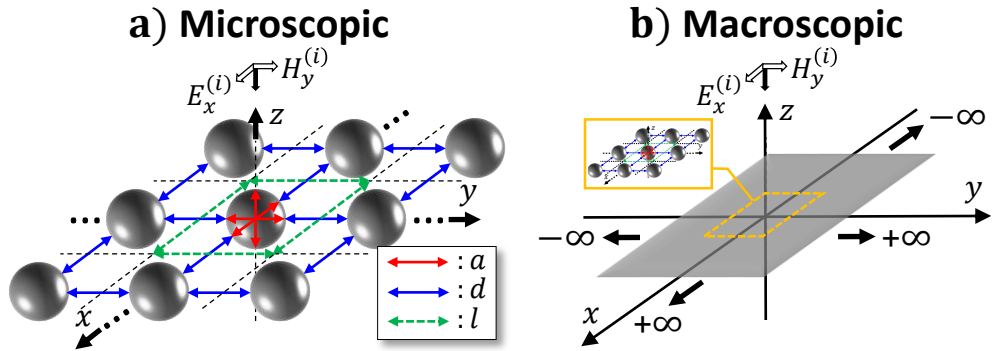

**Figure 1.** Studied metasurface system. The incident light is a plane wave with $E_x^{(i)}$ and $H_y^{(i)}$ components that propagates along the negative $z$ direction. (a) Microscopic view, whereby $a$, $d$, and $l$, are drawn using red, blue, and green arrows, respectively, and denote the diameters of the spheres, gap distances, and the lengths of the period, respectively. (b) Macroscopic view in which we assume that the metasurface is regarded as a uniform plate in the $xy$ plane with an ignorable thickness.

We depict a microscopic picture of the metasurface in Figure 1a where metallic nanospheres with diameters $a$ are periodically arrayed on the $xy$ plane with gap distances $d$, and a period length is equal to $l$. The nanospheres are treated using the jellium model. If the diameter of the nanospheres is sufficiently small, the optical response of the metasurface is characterized by the 2D conductivity defined below. We consider an application of a spatially uniform electric field $\mathbf{E}(t)$ to the metasurface. The electric field is assumed to be linearly polarized in the $x$ direction. The conduction electrons in the spheres are collectively excited and the microscopic electric current density $\mathbf{j}(\mathbf{r}, t)$ appears in the spheres. We discuss below how we evaluate $\mathbf{j}(\mathbf{r}, t)$, either based on quantum mechanics with the use of the TDDFT, or based on the classical theory according to the Drude model.

We introduce a course-grained 2D electric current density expressed as

$$\mathbf{J}_{2D}(t) = \int\int \frac{dxdy}{l^2} \int dz\, \mathbf{j}(\mathbf{r}, t), \qquad (1)$$

where the integrations are carried out over the 2D unit cell area in the $xy$ plane and over the $z$-direction. The $\mathbf{J}_{2D}(t)$ is parallel to the $x$ direction owing to the symmetry of the system. For a sufficiently weak field, the 2D conductivity $\sigma_{2D}(t)$ connects $\mathbf{J}_{2D}(t)$ and the electric field $\mathbf{E}(t)$,

$$\mathbf{J}_{2D}(t) = \int^t dt'\, \sigma_{2D}(t - t')\mathbf{E}(t'). \qquad (2)$$

In the main text, we show quantities in the frequency domain. The quantities in the frequency domain $\mathbf{j}(\mathbf{r}, \omega)$, $\mathbf{J}_{2D}(\omega)$, and $\sigma_{2D}(\omega)$, are given as the time-frequency Fourier transformations of $\mathbf{j}(\mathbf{r}, t)$, $\mathbf{J}_{2D}(t)$, and $\sigma_{2D}(t)$, respectively. For an electric field at a frequency omega, $\mathbf{E}(t) = E_0 \cos(\omega t)$, the electric current density is given by $\boldsymbol{j}(\mathbf{r}, t) = \text{Re}[\boldsymbol{j}(\mathbf{r}, \omega)]\cos(\omega t) - \text{Im}[\boldsymbol{j}(\mathbf{r}, \omega)]\sin(\omega t)$. We refer to $\text{Re}[\boldsymbol{j}(\mathbf{r}, \omega)]$ as the conductive current because it has the same phase relation to the applied electric field, and $\text{Im}[\boldsymbol{j}(\mathbf{r}, \omega)]$ as the dielectric current because it has a phase difference of $\pi/2$.

To elucidate the influence of the electron transport at the gaps, we introduce the following decomposition of the 2D conductivity:

$$\sigma_{2D}(\omega) = \frac{1}{l}\int dx\, \sigma_{2D}(x, \omega). \qquad (3)$$

As is understood from Eq. (2), the conductivity contribution at the gap plane $x = x_G$, $\sigma_{2D}(x_G, \omega)$, which is shown in Fig. 5 of the main text, connects the electric field to the electric current integrated over the gap plane,

$$\int \frac{dy}{l} \int dz\, \mathbf{j}(x_G, y, z, t) = \int dt'\, \sigma_{2D}^{(x)}(x_G, t - t') \mathbf{E}(t'). \tag{4}$$

We now consider a linearly polarized plane wave light irradiation in a direction normal to the metasurface. At the macroscopic scale, we treat the electric current in the metasurface as uniform on the $xy$ plane by ignoring the thickness in the $z$ axis, as shown in Figure 1b. Thus, the macroscopic electric current density is expressed as,

$$\mathbf{J}(\mathbf{r}, t) \approx \delta(z) \mathbf{J}_{2D}(t), \tag{5}$$

where $\mathbf{J}_{2D}(t)$ is the coarse-grained 2D electric current density, which is related to the microscopic current density according to Eq. (1). Substitution of this equation into the macroscopic Maxwell's equations yields the transmission ($T(\omega)$), reflection ($R(\omega)$), and the absorption rates ($A(\omega)$) of the metasurface as follows:[1]

$$T(\omega) = \frac{1}{\left|1 + \frac{2\pi \sigma_{2D}(\omega)}{c}\right|^2}, \tag{6}$$

$$R(\omega) = \left|\frac{\frac{2\pi \sigma_{2D}(\omega)}{c}}{1 + \frac{2\pi \sigma_{2D}(\omega)}{c}}\right|^2, \tag{7}$$

$$A(\omega) = 1 - T(\omega) - R(\omega). \tag{8}$$

Calculated results of $T(\omega)$, $R(\omega)$, and $A(\omega)$ are shown in Figure 3 of the main test.

In practical calculations, we use an impulsive electric field defined by,

$$E_x(\mathbf{r}, t) \approx e_0 \delta(t), \tag{9}$$

where $e_0$ is the amplitude of the impulsive incident field. For this field, the induced current $\mathbf{J}_{2D}(t)$ is proportional to the 2D conductivity $\sigma_{2D}(t)$, and $\sigma_{2D}(\omega)$ can be obtained as the Fourier transform,

$$\sigma_{2D}(\omega) = \frac{1}{e_0} \int dt\, e^{i\omega t} \mathbf{J}_{2D}(t). \tag{10}$$

## 2. TDDFT

For the quantum mechanical description of the electron dynamics in the metasurface, we employ the time-dependent density functional theory (TDDFT) with a jellium model. According to the TDDFT, electron motion is governed by the following time-dependent Kohn-Sham equation with a periodic boundary condition in the $x$ and $y$ directions, and an isolated boundary condition in the $z$ direction,

$$i\hbar \frac{\partial u_{n\mathbf{k}}(\mathbf{r},t)}{\partial t} = \left[\frac{1}{2}\left(-i\hbar\nabla + \hbar\mathbf{k} + \frac{e}{c}\mathbf{A}(t) - e\phi(\mathbf{r},t) + V_{\text{jm}}(\mathbf{r}) + V_{\text{XC}}(\mathbf{r},t)\right)^2\right] u_{n\mathbf{k}}(\mathbf{r},t), \quad (11)$$

where $u_{n\mathbf{k}}$, $\mathbf{k}$, $V_{\text{jm}}$, and $V_{\text{XC}}$, represent the Bloch orbitals, 2D crystalline momentum vector, jellium potential, and exchange correlation potential, respectively. In addition, $\mathbf{A}$ and $\phi$ are the vector and scalar potentials, respectively. In the spherical jellium model, the following charge density $\rho_{\text{jm}}$ with the Wigner-Seitz radius $r_s$ is assumed in the unit cell:[2–4]

$$\rho_{\text{jm}}(\mathbf{r},t) = n^+ \theta\left(\frac{a}{2} - r\right), \quad (12)$$

$$n^+ = \left(\frac{4\pi r_s^3}{3}\right)^{-1}, \quad (13)$$

$$\frac{a}{2} = \left(\frac{3}{4\pi}\frac{N_e}{n}\right)^{\frac{1}{3}}, \quad (14)$$

where $N_e$ is the number of electrons for the metallic nanosphere, and $V_{\text{jm}}$ is the static, 2D periodic ionic potential generated by the background positive charge density $\rho_{\text{jm}}$. The vector potential $\mathbf{A}(t)$ is used to express the spatially uniform electric field $\mathbf{E}(t)$ applied to the metasurface; it is expressed as:

$$\mathbf{A}(t) = -c \int dt\, \mathbf{E}(t). \quad (15)$$

The scalar potential $\phi(\mathbf{r}, t)$ is 2D periodic and satisfies the following Poisson equation with the charge density $\rho(\mathbf{r}, t)$:

$$\Delta\phi(\mathbf{r}, t) = 4\pi\rho(\mathbf{r}, t), \tag{16}$$

$$\rho(\mathbf{r}, t) = -\sum_{n\mathbf{k}} |u_{n\mathbf{k}}(\mathbf{r}, t)|^2, \tag{17}$$

The exchange-correlation potential $V_{\text{XC}}$ is 2D periodic. We assume an adiabatic, local density approximation for it.[5] The microscopic electric current density $\mathbf{j}(\mathbf{r}, t)$ is given by

$$\mathbf{j}(\mathbf{r}, t) = -\text{Re}\left[\sum_{n\mathbf{k}} u_{n\mathbf{k}}^*(\mathbf{r}, t)\left(-i\hbar\nabla + \hbar\mathbf{k} + \frac{e}{c}\mathbf{A}(t)\right)u_{n\mathbf{k}}(\mathbf{r}, t)\right]. \tag{18}$$

Numerical calculations were carried out using SALMON, an open-source code (https://salmon-tddft.jp/) developed in our group.[6] In SALMON, the TDKS equation is solved in real time, using three-dimensional Cartesian grids to express orbitals and potentials. In our calculation, grid spacings $\Delta x$, $\Delta y$, $\Delta z$, and $\Delta t$, are set to 1 Å and $1.25 \times 10^{-3}$ fs, respectively. The time evolution is calculated for duration $T = 12.5$ fs, and the time-frequency Fourier transformations are carried out during this period. The number of $k$ points is chosen from $2 \times 2$ to $6 \times 6$. The convergence with respect to these parameters was carefully examined. The occupation number is determined by the Fermi-Dirac distribution at a temperature of 300 K.

## 3. Classical theory and results

In the classical description, the Drude model is adopted for the microscopic electron dynamics, whereby the electron motion is governed by the Newtonian equation. The equation for the velocity field $\mathbf{v}(\mathbf{r}, t)$ is given by:

$$\frac{\partial \mathbf{v}(\mathbf{r}, t)}{\partial t} = -\gamma\mathbf{v}(\mathbf{r}, t) - \frac{e}{m}\mathbf{E}(\mathbf{r}, t), \tag{19}$$

where $\gamma$ is the collision frequency. The microscopic electric current density $\mathbf{j}_c(\mathbf{r},t)$ in the unit cell, including the origin, is expressed using the velocity field $\mathbf{v}$ as:

$$\mathbf{j}_c(\mathbf{r},t) = -\frac{\omega_p^2}{4\pi e}\theta\left(\frac{a}{2}-r\right)\mathbf{v}(\mathbf{r},t), \qquad (20)$$

where $\omega_p$ is the plasma frequency. In addition, $\gamma$ and $\omega_p$ are chosen to be equal to 5.43 and 0.063 eV, and are determined in such a way that the results of the classical calculations coincide reasonably with those of the TDDFT using the jellium model at $d = 2$ nm. This calculation was carried out again using SALMON. The grid spacings $\Delta x, \Delta y, \Delta z$, and $\Delta t$, were set as 0.25 Å and $4.8 \times 10^{-5}$ fs, respectively.

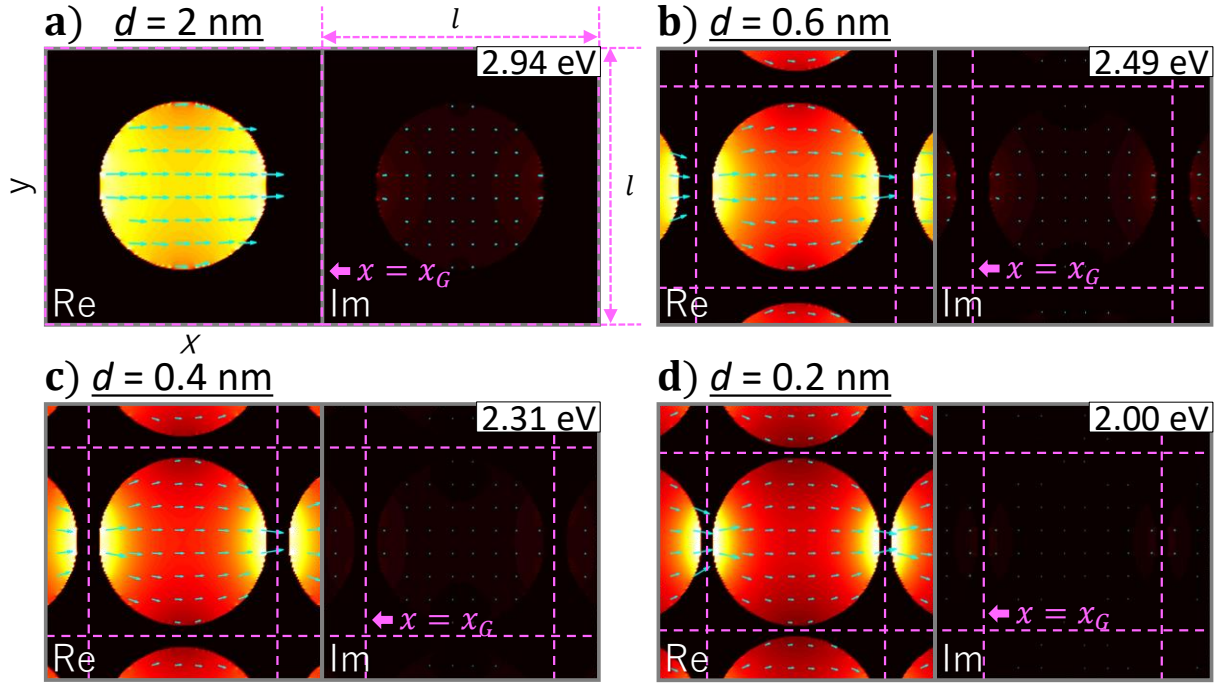

**Figure 2.** (a–d) Spatial distributions of the microscopic electric current density on the $xy$ plane ($z = 0$) at the resonant frequency of the BDP mode, $\omega_{BDP}$, (shown at top right). The maximum intensity is normalized at each figure, and yields $\mathbf{j}_c(x,y,z=0,\omega=\omega_{BDP})/|\mathbf{j}_c|_{\max}$. The real and imaginary parts are shown in the left and the right panels, respectively, for $d = 2, 0.6, 0.4$, and 0.2 nm. Light blue arrows indicate their vector distributions, whereby their lengths are associated with the magnitude. Dashed pink lines indicate the two-dimensional (2D) unit cell with a period $l$. The electron transport takes place through the plane $x = x_G$.

Similar to Figure 4 in the main text, we show the spatial distributions of the microscopic electric current density, $\mathbf{j}_c$, based on the classical theory (Figure 2). Figure 2a–2d shows the results on the $xy$ plane ($z = 0$) at the resonant frequency of each BDP mode, $\omega_{BDP}$. In each figure, the normalized distributions given by $\mathbf{j}_c(x, y, z = 0, \omega = \omega_{BDP})/|\mathbf{j}_c|_{max}$ for $d = 2, 0.6, 0.4, 0.2$ nm are shown. The left and the right panels denote the real and imaginary parts of the current, respectively. Light blue arrows indicate their vector distributions. Dashed pink lines indicate the 2D unit cell with a period $l$. As it has been observed in Figure 4 of the main text, we found an almost spatially uniform current distribution for Re[$\mathbf{j}_c$] at $d = 2$ nm in Figure 2a. However, contrary to the drastic changes from $d = 0.4$ nm to $d = 0.2$ nm in the quantum calculation described in the main text, Figure 2b–d shows almost the same behavior. Accordingly, the currents are more and more localized at both ends of the nanosphere along the $y = 0$ axis as $d$ decreases. Furthermore, for the entire set of values of $d$, Im[$\mathbf{j}_c$] remains vanishingly small. This is simply owing to the fact that the electron transport never takes place across the gap plane $x = x_G$ according to the classical theory when $d > 0$.